\documentclass[pra,english,showpacs,preprintnumbers,amsmath,amssymb,nofootinbib,twocolumn,superscriptaddress]{revtex4-1}

\pdfoutput=1

\usepackage[latin1]{inputenc}
\usepackage{graphicx}
\usepackage{color}
\usepackage{bbm}
\usepackage{amssymb}
\usepackage{amsmath}
\usepackage{tabularx}

\usepackage{dsfont}

\def\0#1#2{\frac{#1}{#2}}

\def\s0#1#2{\mbox{\small{$ \frac{#1}{#2} $}}}

\def\CC{{\mathcal C}}

\newcommand{\beq}{\begin{equation}}
\newcommand{\eeq}{\end{equation}}
\newcommand{\bea}{\begin{eqnarray}}
\newcommand{\eea}{\end{eqnarray}}
\newcommand{\tr}{\mathrm{tr}}

\usepackage{babel}
\makeatother
\begin{document}

\title{Thermodynamics of one-dimensional SU(4) and SU(6) fermions \\ with attractive interactions}

\author{M. D. Hoffman}
\author{A. C. Loheac}
\author{W. J. Porter}
\author{J. E. Drut}

\affiliation{Department of Physics and Astronomy, University of North Carolina, Chapel Hill, North Carolina 27599-3255, USA}

\begin{abstract}
Motivated by advances in the manipulation and detection of ultracold atoms with multiple internal degrees of freedom,
we present a finite-temperature lattice Monte Carlo calculation of the density and pressure equations of
state, as well as Tan's contact, of attractively interacting SU(4)- and SU(6)-symmetric fermion systems in one spatial dimension.
We also furnish a nonperturbative proof of a universal relation whereby quantities computable in the SU(2) case completely determine the
virial coefficients of the SU($N_f$) case.
These one-dimensional systems are appealing because they can be experimentally realized in
highly constrained traps and because of the dominant role played by correlations. The latter are typically
nonperturbative and are crucial for understanding ground states and quantum phase transitions. While quantum
fluctuations are typically overpowered by thermal ones in one and two dimensions at any finite temperature, we find that quantum effects do leave
their imprint in thermodynamic quantities. Our calculations show that the additional degrees of freedom, relative to the SU(2) case, provide
a dramatic enhancement of the density and pressure (in units of their noninteracting counterparts) in a wide region around vanishing
$\beta\mu$, where $\beta$ is the inverse temperature and $\mu$ the chemical potential.
As shown recently in experiments, the thermodynamics we explore here can be measured in a controlled and
precise fashion in highly constrained traps and optical lattices. Our results are a prediction for such experiments in one dimension with
atoms of high nuclear spin.
\end{abstract}

\date{\today}
\maketitle
%%%%%%%%%%%%%%%%%%%%%%%%%%%%%%%%%%%%%%%%%%%%%%%%%%%%%%%%%%%%%
\section{Introduction}

The manipulation and detection of ultracold atoms have recently increased in accuracy and complexity to an extraordinary degree~\cite{RevExp,RevTheory,UltracoldLattices1}.
Along with the realization of atomic microscopes and the trove of possibilities that that entails~\cite{AtomicMicroscopes}, several groups are exploring the nature of
clouds of high-spin atomic species with very stable excited states~\cite{LargeNfReview}, such as alkaline-earth-metal atoms (e.g., Sr) and alkaline-earth-metal-like atoms
(e.g., $^{173}$Yb)~\cite{OrbitalResonancesExp}.
While magnetic Feshbach resonances are absent in those systems (as the total electronic spin is zero), orbital resonances are
available and have recently been shown to be highly controllable with external fields~\cite{OrbitalResonancesTheory}. Those systems were achieved
experimentally in three dimensions, but optical lattices can be tuned to explore their one-dimensional (1D) and two-dimensional (2D) counterparts (see, e.g., Ref.~\cite{1DExp}). Indeed, the 1D case was
first explored relatively recently in Ref.~\cite{1DSUNExp} in the presence of repulsive interactions, where deviations from Luttinger-liquid theory were
observed (see also Ref.~\cite{RMPImambekov}).

Such experimental availability has opened a rather vast set of new possibilities in the form of SU($N_f$)-symmetric systems. Of those, much
is known about the $N_f=2$ case, as revealed by theory and experiment in the last decade; however, much less is known about $N_f > 2$. Indeed,
motivated by the universality of regimes around broad Feshbach resonances, a large amount of work was dedicated to spin-$1/2$
fermions in one, two, and three-dimensions across the BCS-BEC crossover (see, e.g., Refs. ~\cite{RevTheory,UFGBook}). In contrast, theoretical research exploring
the behavior of higher-spin systems has been less common (see, however, Refs.~\cite{HighSpin1,HighSpin2,HighSpin3,HighSpin4} for references on the 1D case).

In this work, we take a step towards quantitatively clarifying the effects of attractive short-range interactions in 1D fermions
with $N_f = 4,6$ internal degrees of freedom (``flavors''). We focus on the thermodynamics and short-range correlations of unpolarized
systems (i.e., every flavor is tuned to the same chemical potential $\mu$). The motivation for 1D systems goes beyond the potential
experimental realization mentioned above. On the theory side, one dimension is interesting because interaction effects are enhanced and
lead to a plethora of collective effects in the form of quasi-long-range order in the ground state, with the accompanying
quantum phase transitions~\cite{Giamarchi,Sachdev}. On the other hand, finite temperature wipes out such transitions, leaving
only traces of interaction effects. The latter, however, can be quantitatively large and theoretically interesting, as we show here.
Furthermore, one dimension is appealing from a methodological perspective for two reasons: first, calculations in one dimension are computationally much
less expensive than in two or three dimensions and thus provide a ``stepping stone'' to higher dimensions that is also physically meaningful;
second, a number of approaches can address 1D systems with contact interactions exactly in the ground state~\cite{TakahashiBook}, but that number
is much reduced at finite temperature~\cite{BetheAnsatzReview}.

Our work focuses on a low-energy effective Hamiltonian of the Gaudin-Yang form~\cite{GaudinYang}
\beq
\label{Eq:H}
\hat H \!=\!\! \int\! dx \!
\left [
\sum_{s}\! \hat \psi_s^\dagger({x})\!\left(\!-\frac{\hbar^2}{2m} \frac{d^2}{dx^2}\!\right)\! \hat \psi_s^{}({x})
- g \! \sum_{s > s'} \hat n^{}_{s}({x}) \hat n^{}_{s'}({x})
\right],
\eeq
where $\hat \psi_s^{\dagger}$ and $\hat \psi_s^{}$ are the creation and annihilation operators in coordinate space for particles of
flavor $s$, and $\hat n^{}_{s} = \hat \psi_s^{\dagger}\hat \psi_s^{}$ are the corresponding density operators.
The sums over $s,s'$ are in the range $1$ to $N_f$, and $g$ is an attractive coupling constant. We examine the cases $N_f = 4,6$, which
represent a continuation of our previous work for $N_f=2$ (see Ref.~\cite{EoS1D}).
Below, we use units such that $\hbar = m = k_B = 1$, where $m$ is the fermion mass.

For the above dynamics, we explore weakly to strongly coupled regimes, as well as a wide range of temperatures.
We accomplish this by putting the system on a lattice in the grand-canonical ensemble and by writing the corresponding
partition function in a field-integral representation, as further explained below. Expectation values of observables are then
estimated using Monte Carlo methods, and we present those results for the particle number density $n$, pressure $P$,
compressibility $\kappa$, and Tan's contact $\mathcal C$~\cite{Tan}.

%%%%%%%%%%%%%%%%%%%%%%%%%%%%%%%%%%%%%%%%%%%%%%%%%%%%%%%%
\section{Many-body method, scales and dimensionless parameters}

We employed the auxiliary-field path-integral Monte Carlo technique, which is now standard in many areas of physics
(see, e.g.,~\cite{Drut:2012md}).
The fermions were placed in a Euclidean space-time lattice of extent $N^{}_x \times N^{}_\tau$ with boundary conditions that are periodic in space and antiperiodic in time.
The physical spatial extent of the lattice is
given by $L = N^{}_x \ell$, where we take $\ell = 1$ and thus set the length and momentum scales.
The temporal lattice is determined by the inverse temperature $\beta = 1/T = \tau N^{}_\tau$, where the time step $\tau= 0.05$
(in lattice units) was chosen to balance discretization effects (see below) and computational efficiency.

A Hubbard-Stratonovich transformation was used to introduce the auxiliary field and thus write the grand-canonical
partition function as a field integral. The latter was evaluated using Metropolis-based Monte Carlo methods, with the sampling
of the auxiliary field carried out using the hybrid Monte Carlo algorithm~\cite{HMC1,HMC2}. As is well known, unpolarized
systems do not suffer from the sign problem as long as the interaction is purely attractive, as is the case here.

The auxiliary-field formalism used here introduces higher-body forces beyond the pairwise interaction that we want to study. If the bare
lattice coupling is $g$, and the temporal lattice spacing is $\tau$, pairwise interactions enter in the path integral at order $A^2 \sim (e^{\tau g} - 1)$;
on the other hand, four-body forces enter at order $A^4$. Our calculations use $\tau=0.05$ and $g < 1.0$ (see below for an explanation of the
physical dimensionless coupling constant $\lambda$), such that $A^2 < 0.05$, and therefore
$A^4 < 0.0025$ in the worst-case scenario. In this fashion, nonuniversal lattice artifacts due to four- and six-body forces were reduced.
No odd-body forces are induced in this formalism.

The physical input parameters are the inverse temperature $\beta$, the chemical potential $\mu$ (the same for all flavors),
and the (attractive) coupling strength $g>0$. From these, we form two dimensionless
quantities: the fugacity and the dimensionless coupling, given by
\beq
z = \exp(\beta \mu)\ \ \ \ \ \text{and} \ \ \ \ \lambda^2 = \beta g^2,
\eeq
respectively.
%In the grand-canonical ensemble, the density $n$ is an output variable, and therefore we
%use $\lambda$ instead of the .
The bare coupling $g$ is simply related to the scattering length $a_0$: $g = 2/a^{}_0$ (see, e.g., Ref.~\cite{ScatteringIn1D}).
Note that $\gamma = g/n$ is often employed in 1D ground-state studies (see, e.g., Refs.~\cite{Tokatly, FuchsRecatiZwerger,RPLD})
as a dimensionless coupling; the form $\lambda^2 = \beta g^2$, however, is more useful at finite temperature because
it encodes the interplay between temperature and interaction effects (i.e., de Broglie wavelength vs size of a two-body molecule).

Lattice Monte Carlo calculations are exact up to systematic (the lattice part) and statistical (the Monte Carlo part) uncertainties.
To address the latter, we took 1000 decorrelated samples for each data point (see plots below), which yields a statistical uncertainty of order $3 \textendash 4\%$.
Controlling the systematic effects amounts to approaching the continuum, infinite-volume limit
while keeping the physics constant (as encoded in the dimensionless parameters $\lambda$ and $z$).
One-dimensional problems enable calculations on large lattices (up to $N_x^{}$ = 141 in this work). For such lattices, the continuum
limit is approached by lowering $\mu$ and increasing $\beta$, which simultaneously ensures that the lattice system is in
the many-particle regime and the thermal wavelength $\lambda^{}_T = \sqrt{2 \pi \beta}$ is in the regime
\beq
1 = \ell \ll \lambda^{}_T \ll L = \ell N^{}_x.
\eeq

Our calculations feature $\lambda^{}_T \simeq 8.0$, which corresponds to $\beta=10$; finite-$\beta$ effects are described in more detail below
(see Appendix \ref{App:Systematics}).
As in our previous study, we verified the approach to the continuum by checking that our results collapse to a universal curve when $\beta$ and $g$
are varied while $\lambda^2=\beta g^2$ is held fixed. Despite the large lattice sizes we used, the systematic finite-$\beta$ effects are apparent for six
flavors at the largest values of $\lambda = 3.0$, as further explained below.

%Because our study proceeded at constant $\lambda$, increasing $\beta$ implies
%reducing $g$, which results in smaller uncertainties associated with the temporal lattice spacing $\tau$ in the
%Trotter-Suzuki decomposition; these are expected to be of order $1\!-\!2\%$ (see e.g. Ref.~\cite{BDM2}).

%%%%%%%%%%%%%%%%%%%%%%%%%%%%%%%%%%%%%%%%%%%%%%%%%%%%%%%%
%%%%%%%%%%%%%%%%%%%%%%%%%%%%%%%%%%%%%%%%%%%%%%%%%%%%%%%%

\section{Results}

In this section we present our numerical results on the thermodynamics of $N_f$-flavor fermions with
attractive interactions for $N_f=4, 6$, along with
a universal relation whereby the dynamics of the two-flavor problem determines the virial coefficients of the $N_f$-flavor case.
Our results are shown in dimensionless form as a ratio of a physical quantity and its noninteracting counterpart, both evaluated at
identical input parameters. In some instances, this was accomplished by scaling the appropriate power of the thermal wavelength
$\lambda^{}_T = \sqrt{2 \pi \beta}$.

As advertised above, our main result, as a direct output of our lattice calculations, is the density equation of state
$n(\lambda, \beta \mu, N^{}_f)$. From that function we obtain the pressure $P(\lambda, \beta \mu, N^{}_f)$ by integration with
respect to $\beta \mu$ and the isothermal compressibility by differentiation with respect to the same parameter.
In addition, we present Monte Carlo results for Tan's contact $\CC$ which we obtained by relating it to the interaction energy.

%%%%%%%%%%%%%%%%%%%%%%%%%%%%%%%%%%%%
\subsection{Density}
\begin{figure}[t]
\includegraphics[width=1.0\columnwidth]{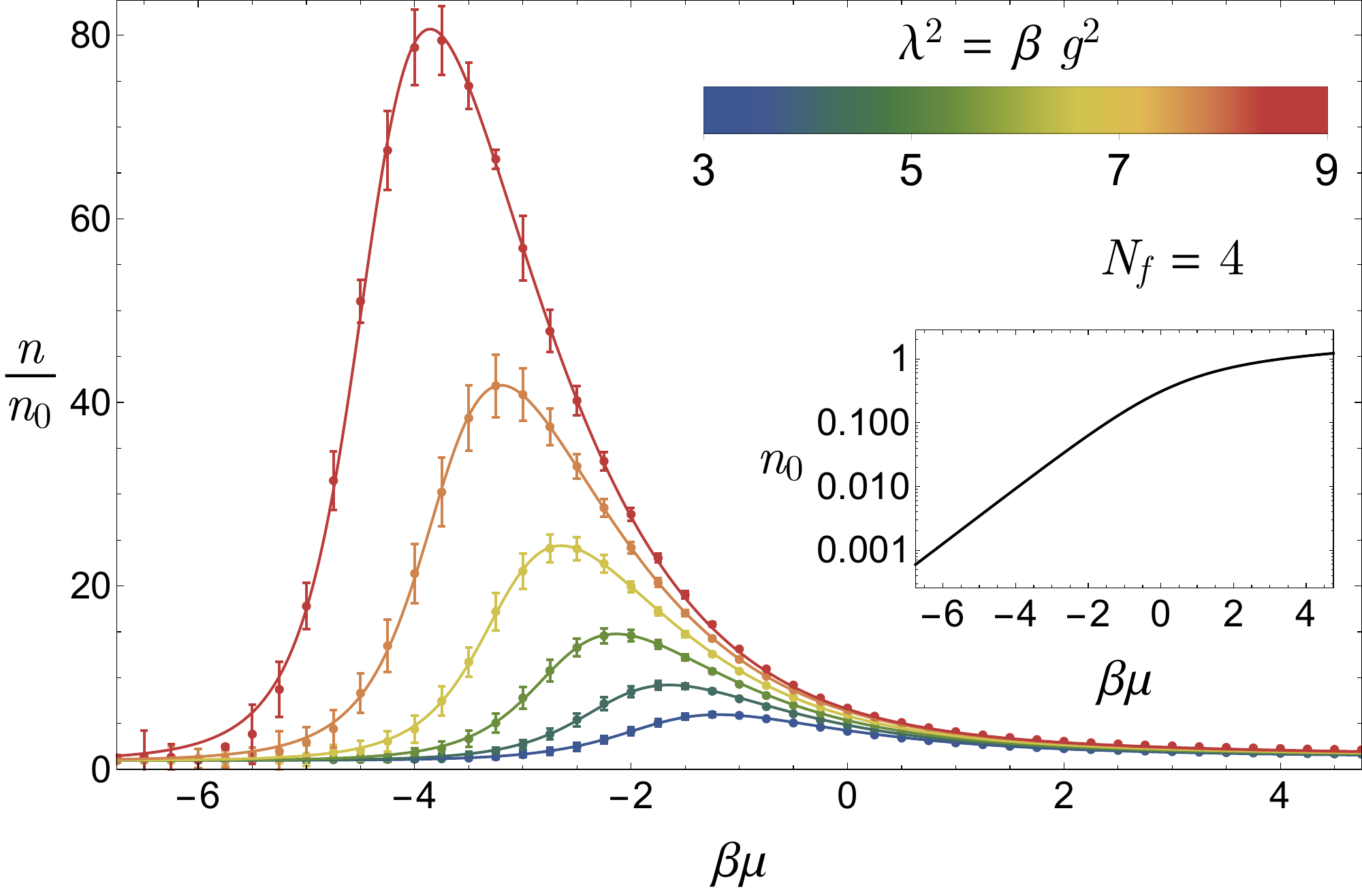}
\includegraphics[width=1.0\columnwidth]{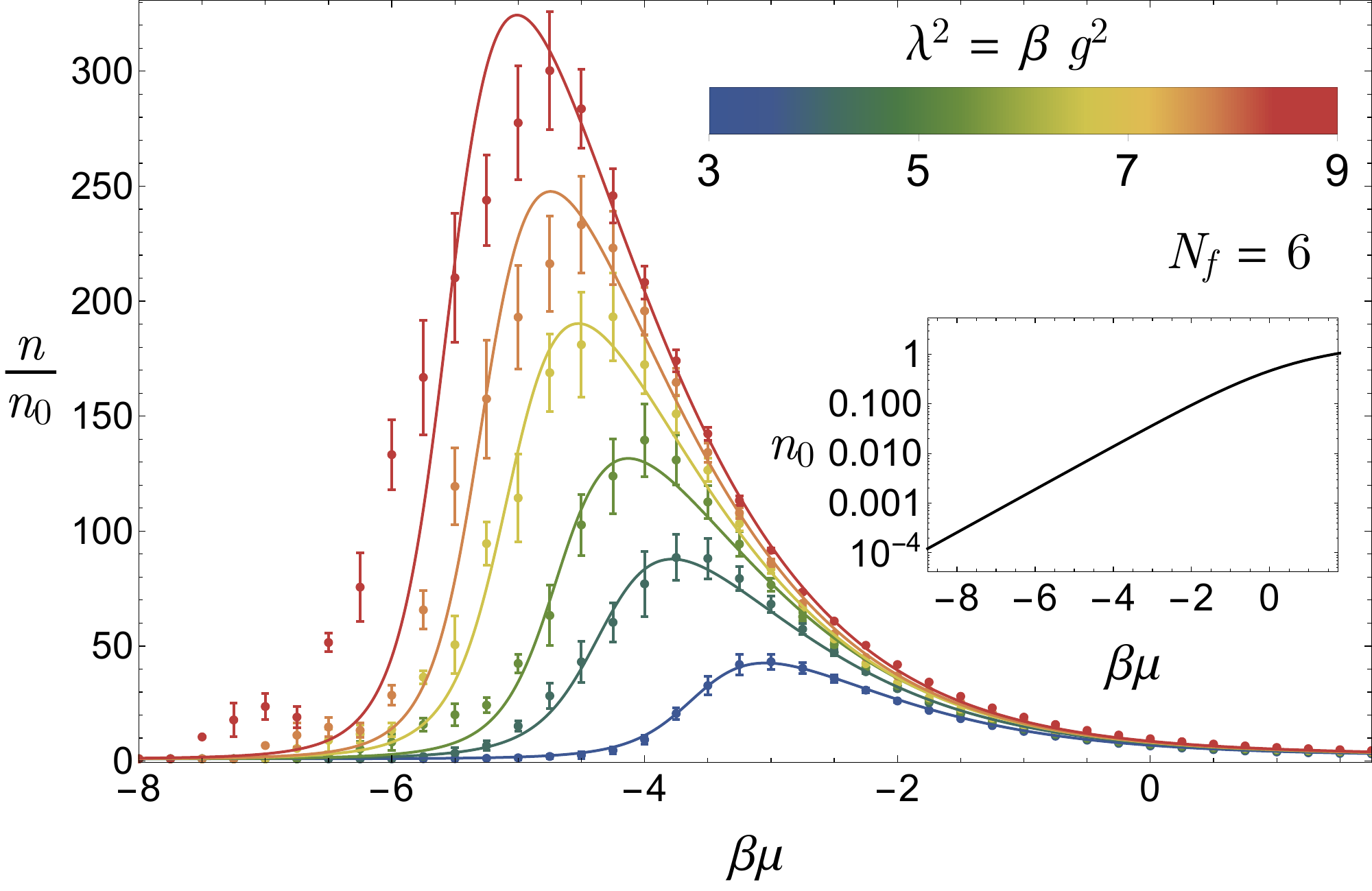}
\caption{\label{Fig:n_n0_nf4}(Color online) Density $n$ for $N^{}_f=4$ (top) and $N^{}_f=6$ (bottom), in units of the density of the noninteracting system $n^{}_0$ (inset), as a function of the dimensionless parameters $\beta\mu\!=\!\ln z$ and $\lambda^2\!=\!\beta g^2$. From bottom to top, the coupling is $\lambda\!=\!1.75, 2.0, 2.25, 2.5, 2.75, 3.0$. The data points come from the quantum Monte Carlo calculations and the solid lines are from the fits (see Eq.~\ref{Eq:fit}).
}
\end{figure}
%%%%%%%%%%%%%
In Fig.~\ref{Fig:n_n0_nf4} we present the density $n$ for $N_f=4,6$ respectively, in units of the noninteracting density $n_0$, as a function of the dimensionless parameters $z$ and $\lambda$, defined above. The noninteracting result is
\beq
\label{Eq:n0}
n^{}_0 \lambda^{}_T = \frac{N_f}{\sqrt{\pi}} I^{}_1(z),
\eeq
where $I^{}_1(z) = z\,{d I^{}_0(z)}/{d z}$, and
\beq
I^{}_0(z) = \int_{-\infty}^{\infty}dx \ln\left (1 + z e^{-x^2} \right).
\eeq
As is well known, one may write these integrals in terms of polylogarithms: $I_0(z) = -\sqrt{\pi} \text{Li}_{3/2}(-z)$ and $I_1(z) = -\sqrt{\pi}
\text{Li}_{1/2}(-z)$, where $\text{Li}_s$ is the polylogarithm function of order $s$.

The solid curves in Fig.~\ref{Fig:n_n0_nf4} correspond to an empirical fit determined from the original Monte Carlo data, as given by
Eq.~(\ref{Eq:fit}) below. The error bars are given by the standard deviation of the density operator in the Monte Carlo data.
For each $\lambda >0$ there exists a strongly coupled
regime around a negative value of $\beta \mu = \ln z$, where the deviation from the noninteracting answer is maximal. The maxima can be shown to satisfy
$n_0 \kappa_0 = n \kappa$, where $\kappa$ is the isothermal compressibility of the system at finite $\lambda$, and $\kappa_0$ is the noninteracting result. This relation can be easily seen by setting
\beq
\frac{\partial (n/n_0)}{\partial \mu} = 0,
\eeq
and using the definition of $\kappa$ of Eq.~(\ref{Eq:KappaDef}).

These results are qualitatively very similar to those of our previous work of Ref.~\cite{EoS1D} for the two-flavor system. The effects of interactions are clearly enhanced by increasing the number of flavors. In general, the regions with the largest departure from noninteracting results is larger and shifts lower in $\beta \mu$ with increasing $\lambda$ or increasing $N_f$.

%%%%%%%%%%%%%%%%%%%%%%%%%%%%%%%%%%%%
\subsection{Pressure and compressibility}
\begin{figure}[t]
\includegraphics[width=1.0\columnwidth]{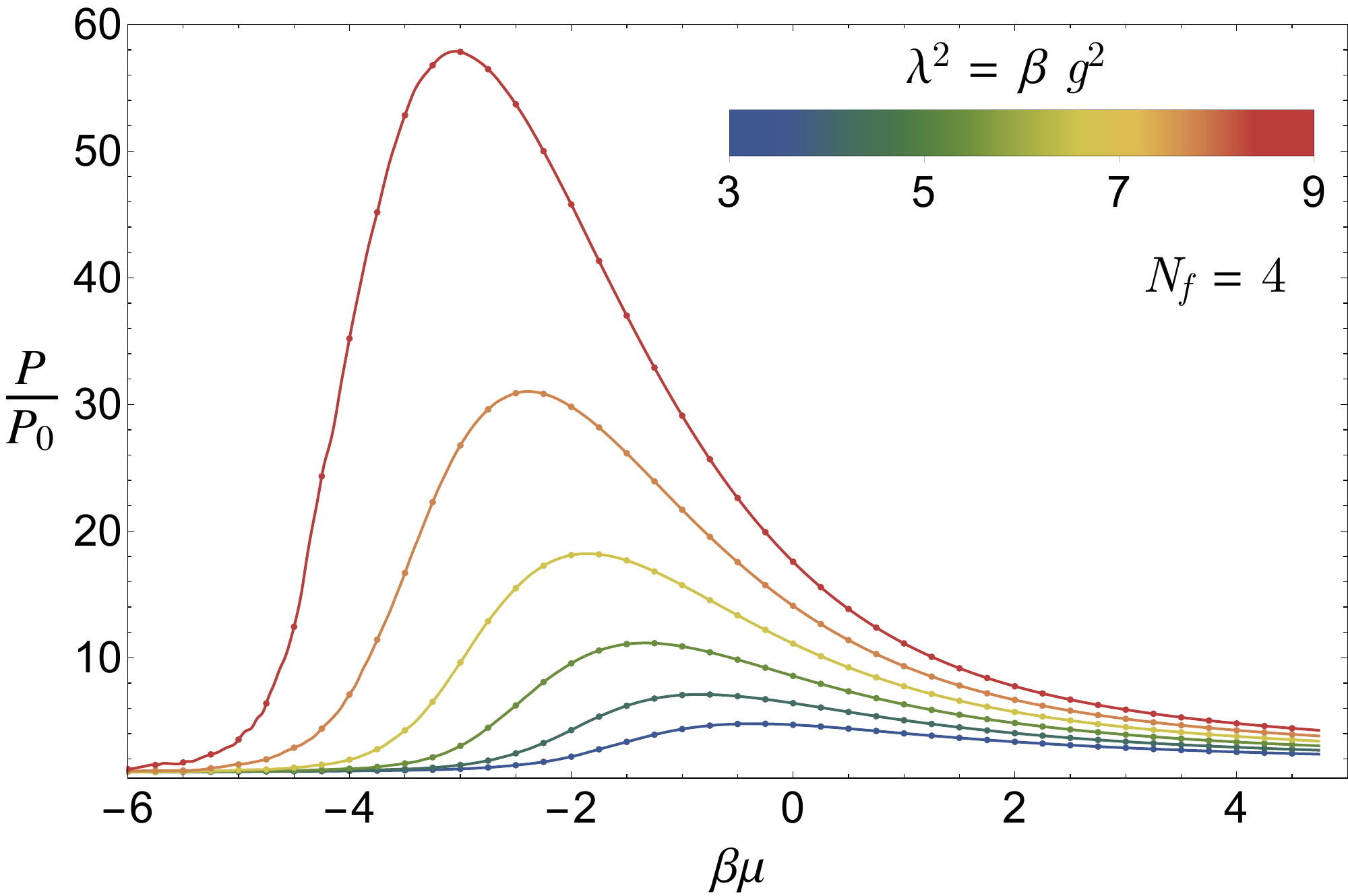}
\includegraphics[width=1.0\columnwidth]{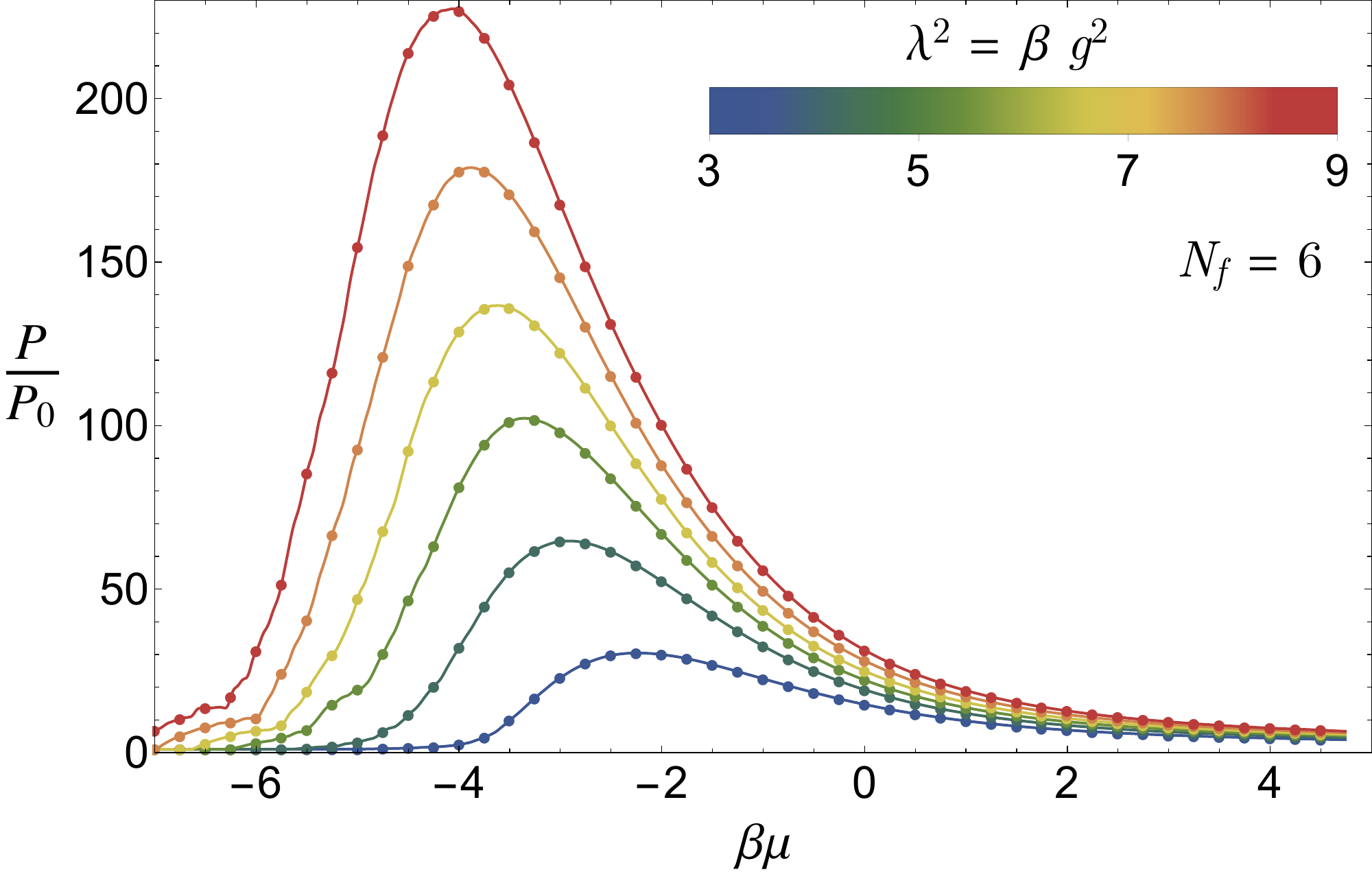}
\caption{\label{Fig:P_P0_nf4}(Color online) Pressure for $N_f = 4$ (top) and $N_f = 6$ (bottom) in units of its noninteracting counterpart, as a function of the dimensionless parameters $\beta \mu = \ln z$ and $\lambda^2 = \beta g^2$, obtained by $\beta\mu$ integration of the density (see Eq.~\ref{Eq:PIntegration}). The values of $\lambda$ shown in this plot are the same as in Fig.~\ref{Fig:n_n0_nf4}.}
\end{figure}

\begin{figure}[t]
\includegraphics[width=1.0\columnwidth]{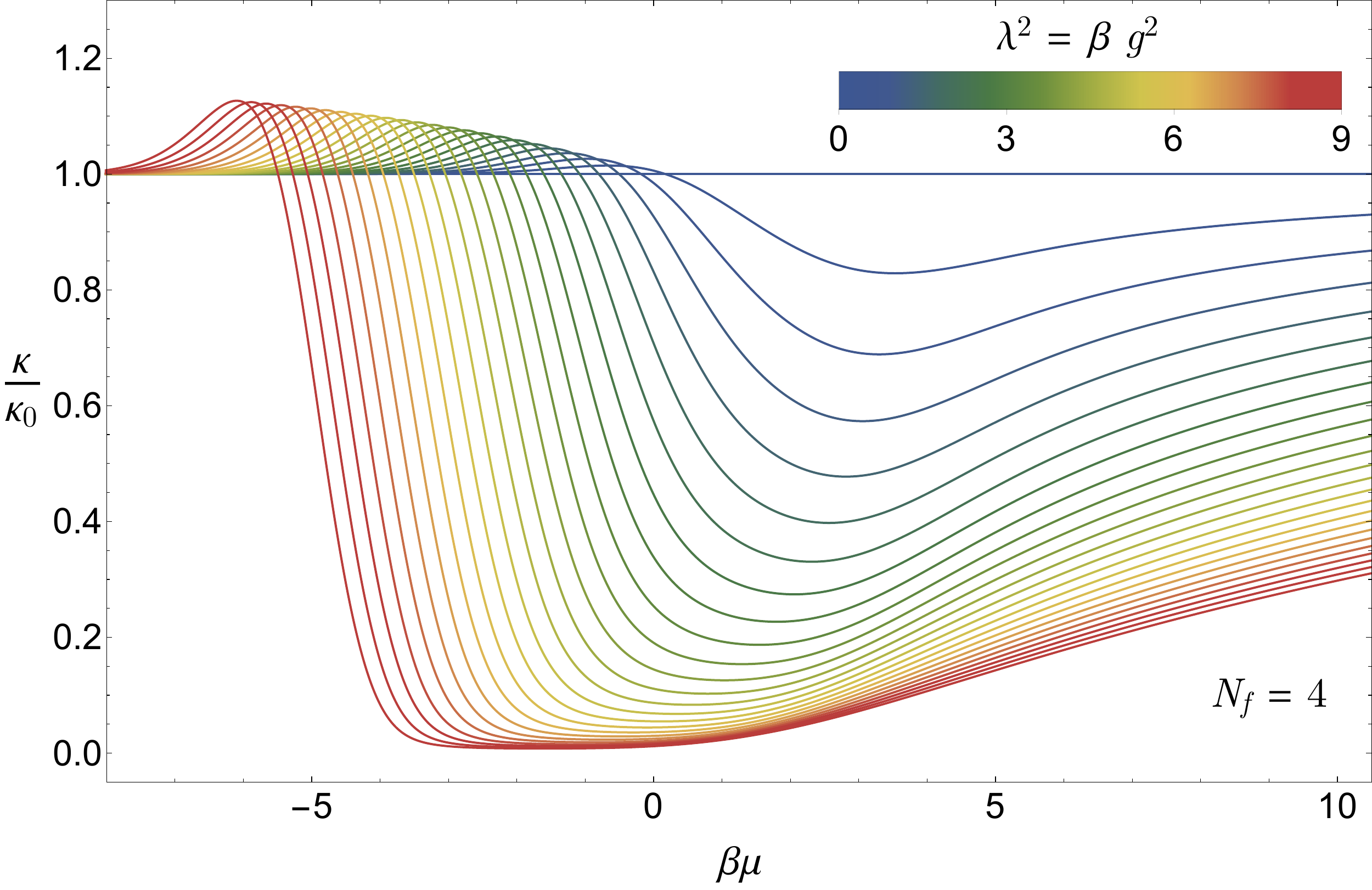}
\includegraphics[width=1.0\columnwidth]{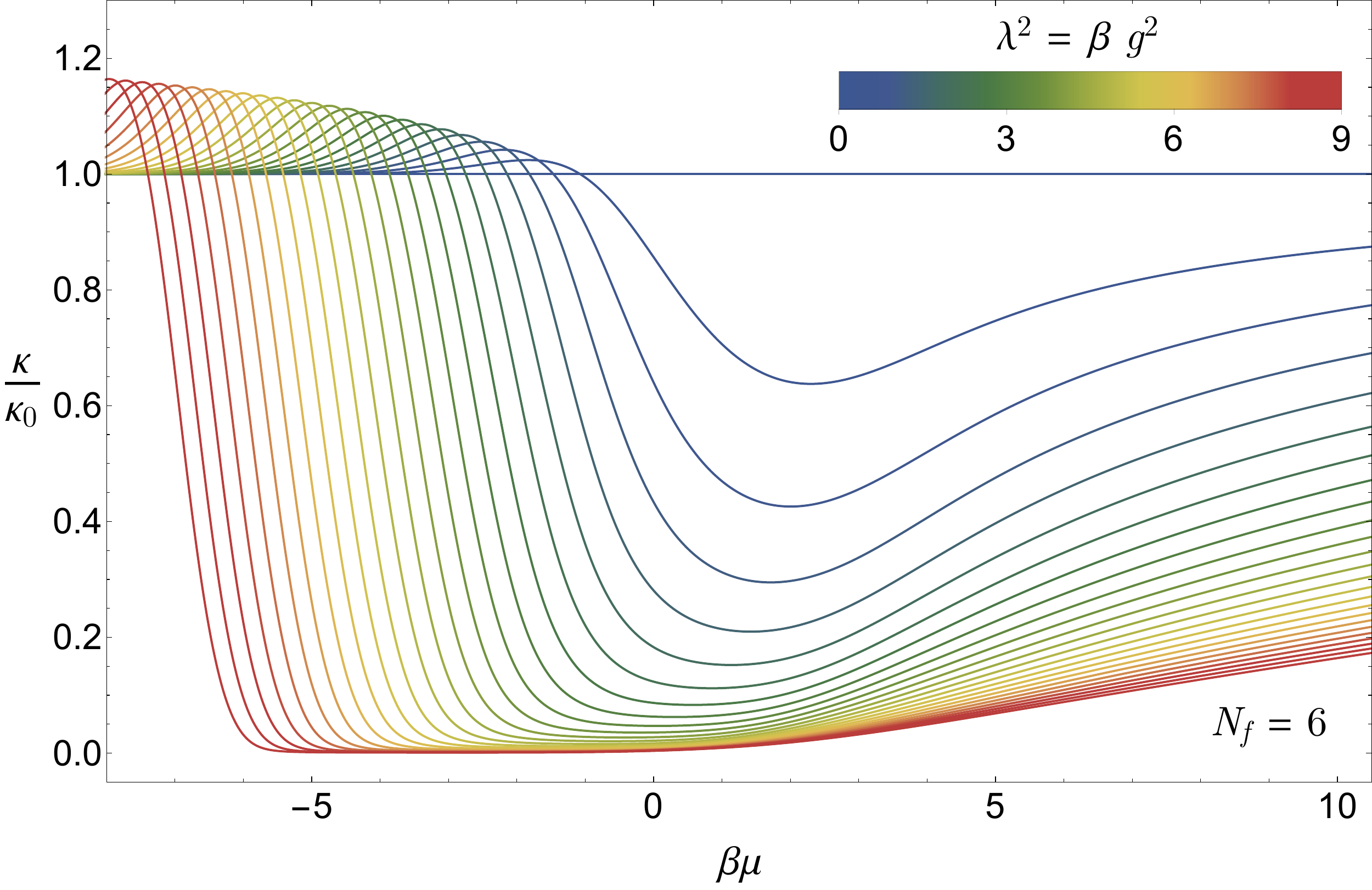}
\caption{\label{Fig:kappa_nf4}(Color online) Isothermal compressibility for $N_f = 4$ (top) and $N_f = 6$ (bottom) in units of its noninteracting counterpart, as a function of the dimensionless parameters $\beta \mu = \ln z$ and $\lambda^2 = \beta g^2$. The values of $\lambda$ range from 0 to 3.0 in steps of 0.125.}
\end{figure}

We estimate the pressure by integrating $n \lambda^{}_T$
over $\log z = \beta \mu$. We use the $z=0$ limit (i.e., $\beta \mu \to -\infty$) as a reference point; therefore, we verify that the data tend (within statistical uncertainties) to
the virial expansion at low $z$ and use
that result at second order to complete the integration to $z=0$. The second-order virial coefficient can be obtained from its value for $N_f=2$ (see below).
In this limit the pressure vanishes, so that
\beq
\label{Eq:PIntegration}
P\lambda_T^3 = 2\pi \int_{-\infty}^{\beta \mu} {n \lambda_T} \; d (\beta \mu)'.
\eeq
The results for $P$, in units of the noninteracting pressure $P_0$, are shown in Fig.~\ref{Fig:P_P0_nf4}. The free gas pressure is given by
\beq
P^{}_0\lambda_T^3 = 2 N_f \sqrt{\pi} I^{}_0(z),
\eeq
where $I^{}_0(z)$ is given above.
The derivative of the density $n$ yields the isothermal compressibility,
\beq
\label{Eq:KappaDef}
\kappa = \frac{\beta}{n^2}\left . \frac{\partial n}{\partial (\beta \mu)} \right |^{}_\beta =
\lambda^{3}_T \frac{\sqrt{2\pi}}{(n \lambda^{}_T)^2} \left . \frac{\partial (n \lambda^{}_T)}{\partial (\beta \mu)} \right |^{}_\beta .
\eeq
We show this quantity in Fig.~\ref{Fig:kappa_nf4}, in units of the respective noninteracting counterpart $\kappa^{}_0$, where (in dimensionless form)
\beq
\kappa^{}_0 \lambda_T^{-3} = N_f \pi^{-3/2} (n^{}_0 \lambda^{}_T)^{-2} I_2(z),
\eeq
and $I^{}_2(z) = z\,{d I^{}_1(z)}/{d z}$. These plots were generated by taking a derivative of the fits to the density data.

As expected, in the limits of large $\beta \mu$ (both positive and negative), $\kappa \to \kappa^{}_0$.
The attractive interaction, combined with Pauli exclusion, gives rise to hard-core bosonic molecules at strong coupling,
which makes the system much less compressible in that region, which in turn yields $\kappa \ll \kappa^{}_0$ there.
Indeed, weaker couplings are much less affected by such hard-core binding.

%%%%%%%%%%%%%%%%%%%%%%%%%%%%%%%%%%%%%%%
\subsection{Tan's contact}

To determine Tan's contact, we rely on the expectation value of the interaction energy $\langle \hat V \rangle$.
By definition,
\beq
{\mathcal C} = \frac{2}{\beta \lambda^{}_T}\left . \frac{\partial (\beta \Omega)}{\partial (a^{}_{0}/\lambda^{}_T)} \right |^{}_{\mu,T},
\eeq
where $\Omega$ is the grand thermodynamic potential. Using the Feynman-Hellman theorem on the grand-canonical partition function, we obtain
\beq
{\mathcal C} = -{g} {\langle \hat V \rangle}. % = -\frac{\sqrt{2\pi} \lambda}{\lambda^{}_T} {\langle \hat V \rangle}.
\eeq
Note that ${\mathcal C}$ can be made dimensionless and intensive by multiplying it by $\lambda^{4}_T/L$.

%%%%%%%%%%%%%
\begin{figure}[h]
\includegraphics[width=1.0\columnwidth]{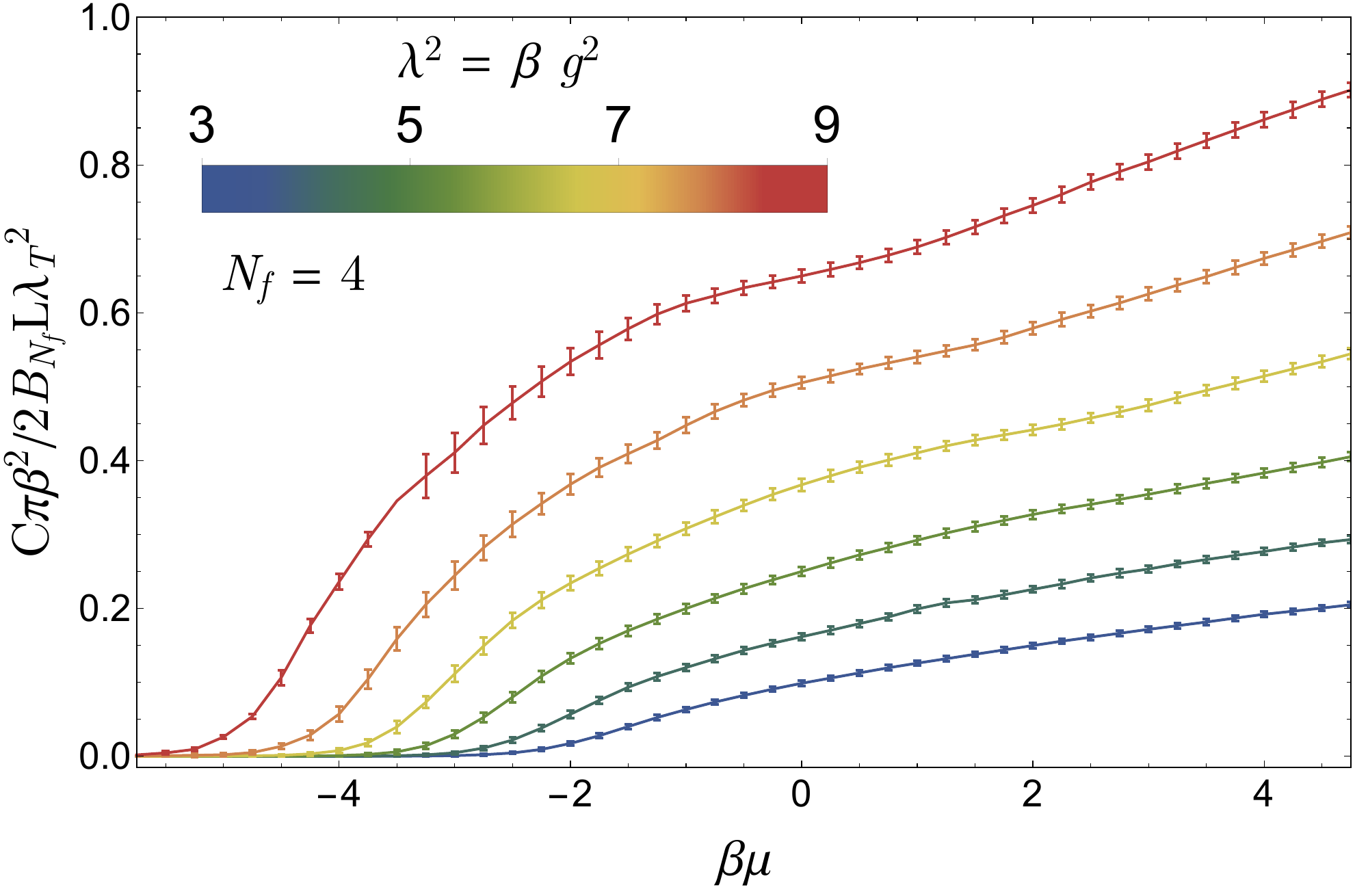}
\includegraphics[width=1.0\columnwidth]{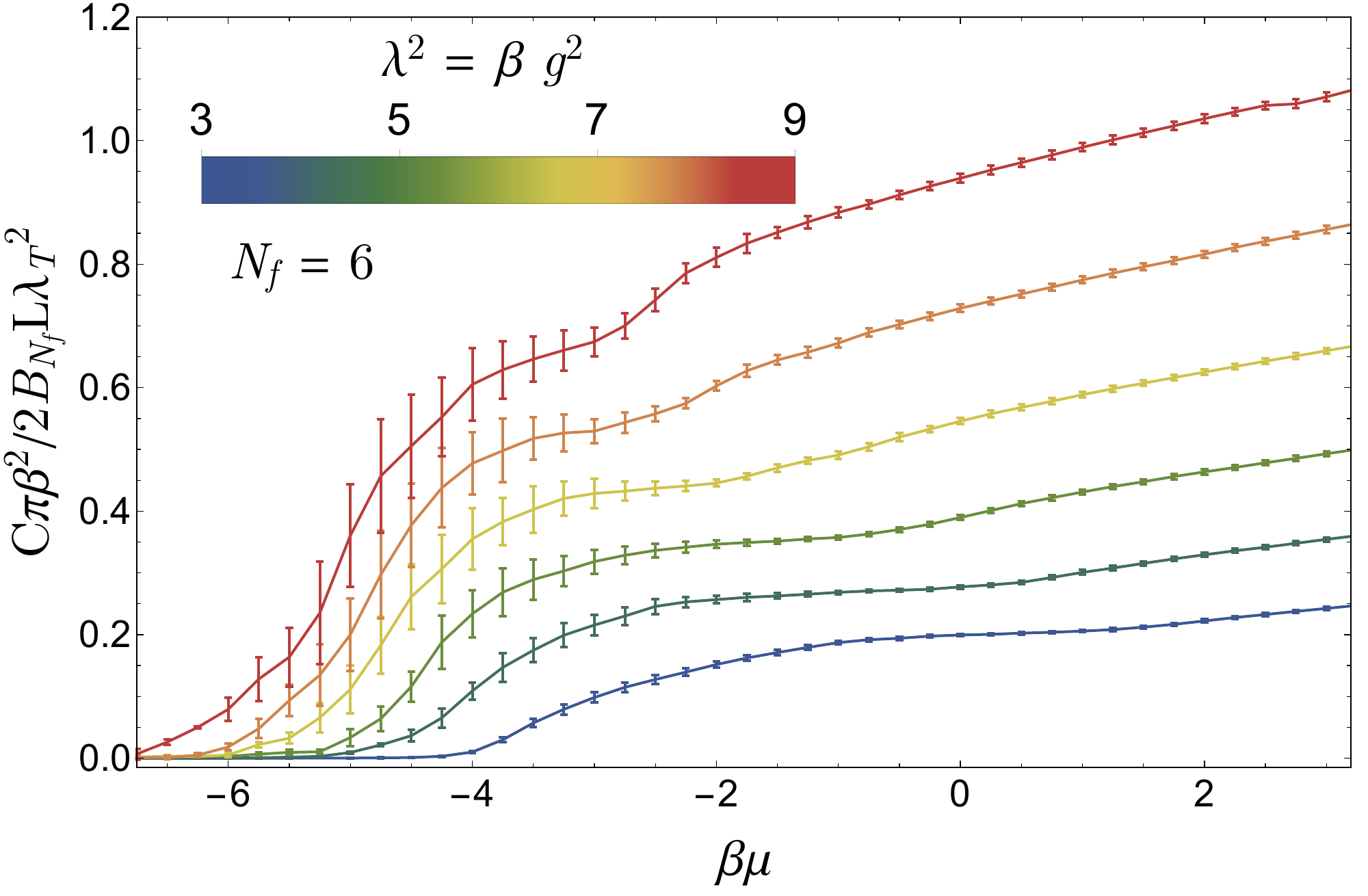}
\caption{\label{Fig:Contact_nf4}(Color online) Tan's contact $\mathcal C$ for $N_f=4$ (top) and $N_f=6$ (bottom), scaled by
$\beta \lambda^{}_T/(2 Q^{}_1 \lambda^2 )  = \pi \beta^2/(2 B_{N_f} L
\lambda^2 )$, as in Ref.~\cite{EoS1D}, as a function
of $\beta \mu$, for $\lambda = 1.75, 2.0, 2.25, 2.5, 2.75, 3.0$, which appear from bottom to top. The value $B_{N_f}$ is the binomial coefficient $N_f$ choose 2; this scale factor was chosen to facilitate  comparison between flavors.}
\end{figure}

In Fig.~\ref{Fig:Contact_nf4} we show our results for the contact. The size of the statistical error bars and the smoothness of
the central values show that statistical effects are generally well controlled across $\beta \mu$. The systematic effects,
on the other hand, are likely larger for strong coupling than for weak coupling (see discussion of systematics below).
As in our previous paper, we note that both the $N_f = 4$ and $N_f = 6$ data for the contact become approximately linear in $\beta \mu$
for $\beta \mu \geq 1.5$. In that regime, the contact satisfies
\beq
\mathcal C \pi \beta^2/(2 L \lambda^2 ) = \langle \hat n^{}_\downarrow \hat n^{}_\uparrow \rangle \pi \beta/2 \to \zeta^{}_1 \beta \mu + \zeta^{}_2,
\eeq
where we find $\zeta^{}_1 = 0.21(1)$ for $N_f = 4$ and $\zeta^{}_1 = 0.48(1)$ for $N_f = 6$. As in the $N_f = 2$ case, density-density correlations in the noninteracting gas
leave an imprint at all couplings. As is evident from the plot, $\zeta^{}_2(\lambda) \simeq a + b \lambda$ is approximately linear
in $\lambda$ at large $\beta \mu$. In the $N_f =4$ case $a = 1.0(1)$ and
$b = 0.5(4)$; the values for $N_f = 6$ are $a = 2.9(1)$ and
$b = 2.2(4)$ when extrapolated to $\beta\mu=10$.
%Analytic estimates in the absence of interactions yield
%$\zeta^{}_1 = 1/\pi = 0.318...$ and $\zeta^{}_2 \propto (\beta\mu)^{-1}$.
%Although much is known about $\mathcal C$ in various situations (see e.g. Ref.~\cite{ContactReview} for a review),
%the full temperature dependence in 1D shown here does not appear anywhere else in the literature, to the best of our knowledge.
%

%%%%%%%%%%%%%%%%%%%%%%%%%%%%%%%%%%%%%%%%%%%%%%%%%%%%%%
\subsection{The virial expansion and a universal relation for virial coefficients across different $N_f$}

In high-temperature dilute regimes where $z \ll 1$, the virial expansion can be a very useful approximation.
Recent years have, in fact, seen a resurgence of interest in the calculation of progressively higher-order
virial coefficients $b_n$ either by exact diagonalization of the few-body problem (see, e.g., Ref. ~\cite{VirialCoefficientED}) or by designing \emph{ad hoc}
Monte Carlo methods~\cite{VirialCoefficientMC}.

Here we show that the $b_n$ for the $N_f$-flavor system are determined by the $N_f=2$ problem.
This property may be intuitively anticipated, as the physics is set entirely by pairwise interactions.
However, the proof itself is enlightening and we therefore show it here in some detail.

We begin by stating more explicitly the form of the field-integral representation of the grand-canonical partition function,
which is given by
\beq
\mathcal Z \equiv \text{Tr}\left[e^{-\beta (\hat H - \mu \hat N)} \right] =
\int \mathcal D\sigma \ {\det}^{N_f}{(1+z\, {\mathcal U}[\sigma])},
\eeq
where, as before, $\beta$ is the inverse temperature and $z$ is the fugacity. The field $\sigma$ is an
auxiliary Hubbard-Stratonovich scalar and the matrix ${\mathcal U}[\sigma]$ encodes the dynamics of
the system. The precise form of ${\mathcal U}[\sigma]$ is not important for the derivations that follow, in the sense that
it applies to completely general two-body interactions (not just point like), which reflects the universality of the result
(see Appendix~\ref{App:Derivation} for a schematic derivation of the form of $\mathcal Z$ for $N_f$ flavors; further details
can be found in the literature: see, e.g., Ref.~\cite{Drut:2012md}).

The virial coefficients are defined by
\beq
\label{eq:VirialBderivativeDef}
b_m^{} = \frac{1}{Q^{}_1}\frac{1}{m!}\left . \frac{\partial^m \ln \mathcal Z}{\partial z^m} \right|_{z=0} ,
\eeq
where $Q_1^{} = N_f L / \lambda^{}_T$ is the single-particle partition function.
We next consider the cumulant expansion of $\ln \mathcal Z$, which reads
\beq
\label{eq:CumulantExpansionZ}
\ln \mathcal Z = \sum_{n=1}^{\infty} \frac{\kappa_n[Y,N_f]}{n!} ,
\eeq
where $\kappa_n[Y,N_f]$ are the cumulants of
\beq
Y(\sigma; z) = \ln{\det}^{N_f}{(1\!+\!z\,{\mathcal U}[\sigma])} = N_f \ln{\det}{(1\!+\!z\,{\mathcal U}[\sigma])}.
\eeq
For $N_f$ even, $Y(\sigma; z)$ is real by definition, and we can make that explicit by writing, instead of the above,
\beq
Y(\sigma; z) = \frac{N_f}{2} \ln(|{\det}{(1\!+\!z\,{\mathcal U}[\sigma])}|^2).
\eeq
For $N_f$ odd, on the other hand, we must account for the fact that $N_f$ does not eliminate the sign of the determinant,
which results in an imaginary part for $Y(\sigma; z)$:
\beq
Y(\sigma; z) = \frac{N_f}{2} \left\{ \ln(|{\det}{(1\!+\!z\,{\mathcal U}[\sigma])}|^2) + 2 i \theta[\sigma]\right\},
\eeq
where $\theta$ can take on the values zero or $\pi$ if the determinant is purely real, as in the cases considered here.
Note that these assumptions may be relaxed to some extent: As long as
the system is balanced (in mass and spin), such that different flavors
are otherwise identical, the determinant in $\mathcal Z$ will appear raised to
the power of $N_f$, such that the above derivations are essentially unchanged.
However, it should be borne in mind that such generalizations make the
determinant complex for repulsive interactions, such that the phase angle $\theta$
plays a crucial role in those cases.

The cumulants obey the usual definition, namely,
\bea
\kappa_1 &=& \langle Y \rangle, \\
\kappa_2 &=& \langle Y^2 \rangle - \langle Y \rangle^2, \\
\vdots \nonumber
\eea
and so on, where $\langle \cdot \rangle$ denotes the path-integral expectation value over $\sigma$ with unit measure.
Clearly, the $\kappa_n[Y,N_f]$ contain all the information about the dynamics of the system, although it is \emph{a priori} unknown whether the
expansion even converges. As long as the thermodynamics of the system [i.e., the left-hand side of Eq.~(\ref{eq:CumulantExpansionZ})]
is well defined, however, the sum makes sense at least formally.

The role of the phase fluctuations for odd $N_f$ can be seen more explicitly by separating $Y$ into its real and
imaginary parts, $Y = Y_R + iY_I$, and writing the cumulants in terms of those:
\bea
\kappa_1 &=& \langle Y_R \rangle + i \langle Y_I \rangle, \nonumber \\
\kappa_2 &=& \langle Y_R^2 \rangle - \langle Y_R \rangle^2 - (\langle Y_I^2\rangle - \langle Y_I \rangle^2 ) \nonumber \\
&&\ \ \ \  +\ 2i (\langle Y_R Y_I \rangle - \langle Y_R \rangle \langle Y_I \rangle ) , \nonumber  \\
\vdots . \nonumber
\eea
The imaginary part of the cumulants should add up to zero in the full sum of Eq.~(\ref{eq:CumulantExpansionZ}),
because we know $\ln \mathcal Z$ is a real quantity. Therefore, the imaginary part of the cumulants plays no role
and can be safely ignored. However, we see from the above that $Y_I$ itself does enter in the real part of $\kappa_n$
for $n>2$, and it does so in a well-defined way through the properties of the distribution of the phase angle $\theta$.
Such distributions have been the source of much discussion in the context of lattice QCD at finite chemical potential
(see e.g. Refs.~\cite{QCDFiniteMu1,QCDFiniteMu2}) and have also been recently explored in nonrelativistic systems~\cite{NoiseWJPJED}.

In both the even- and odd-$N_f$ cases, the above cumulants $\kappa_n$ satisfy a homogeneity property whereby
\beq
\kappa_n[Y,N_f] =  \left (\frac{N_f}{2}\right)^n \kappa_n[Y,2].
\eeq
Putting together Eqs.~(\ref{eq:VirialBderivativeDef}) and~(\ref{eq:CumulantExpansionZ}), along with the
homogeneity property, shows that the thermodynamics of SU($N_f$) systems is governed by
quantities that can be computed entirely within the SU($2$) theory. (Note that if $N_f$ is odd, one must account
for the sign of the determinant, even if the SU($2$) theory has no information about it.)
In particular, homogeneity allows us to analyze the relationship between virial expansions across different values of $N_f$.
Indeed, it is easy to see that the leading order is
\beq
\left.\frac{\partial \kappa_1[Y,N_f]} {\partial z} \right|_{z=0} = N_f \langle \tr\ \mathcal U[\sigma] \rangle =  Q_1^{}
\eeq
and
\beq
\left.\frac{\partial \kappa_n[Y,N_f]} {\partial z} \right|_{z=0} = 0
\eeq
for all $n > 1$. This is, of course, consistent with the fact that $b^{}_1 = 1$ by definition.
Moreover, all $m$th derivatives for $m < n$ vanish upon evaluation at $z = 0$, such that
the expressions for the $b_m$ in terms of the $\kappa_n$ contain a finite and small number of terms:
\bea
b_2^{} &=& \frac{1}{Q^{}_1}\frac{1}{2!}
\left . \left [
 \frac{\partial^2 \kappa_1} {\partial z^2}  +
\frac{1}{2!} \frac{\partial^2 \kappa_2} {\partial z^2}
\right]  \right|_{z=0} , \\
b_3^{} &=& \frac{1}{Q^{}_1}\frac{1}{3!}
\left . \left [
 \frac{\partial^3 \kappa_1} {\partial z^3}  +
\frac{1}{2!} \frac{\partial^3 \kappa_2} {\partial z^3}  +
\frac{1}{3!} \frac{\partial^3 \kappa_3} {\partial z^3}
\right] \right|_{z=0} , \nonumber \\
&\vdots&
\eea
and so on. The above is valid for any $N_f$ and can be summarized as
\beq
\label{eq:virialandcumulantfinal}
b_m^{}(N_f) =\frac{1}{Q^{}_1} \frac{1}{m!}\sum_{n=1}^{m} \left . \frac{1}{n!} \frac{\partial^m \kappa_n[Y,N_f]} {\partial z^m} \right|_{z=0} .
\eeq
Thus, using the cumulant property mentioned above,
\beq
\label{eq:virialandcumulantNfvs2}
b_m^{}(N_f) =\frac{1}{Q^{}_1} \frac{1}{m!}\sum_{n=1}^{m} \left . \frac{N_f^n}{2^n n!} \frac{\partial^m \kappa_n[Y,2]} {\partial z^m} \right|_{z=0} .
\eeq

Equation~(\ref{eq:virialandcumulantNfvs2}) shows the anticipated result, namely, that the virial coefficients of the $N_f$-flavor
system are fully determined quantities that can be computed in the two-flavor case; the crucial quantities are the derivatives of
the $\kappa_n$ cumulants. The latter can of course be written in terms of canonical partition functions, which leads to the well-known
expressions for the virial coefficients.

We stress that the above connection between the general $N_f$ and $N_f=2$ field theories does not imply a simple relationship between
virial coefficients across theories. This is immediately apparent in the $N_f=3$ case (see our comment on odd $N_f$ below), which displays the Efimov effect and is thus
fundamentally different from the $N_f=2$ case. Our proof simply states that the underlying quantities determining the $b_n$ (i.e., the cumulants
and their derivatives) are the same for all theories and can be computed at $N_f=2$. The relationship cannot be inverted to yield an
equation for $b_n$ at arbitrary $N_f$ as a function of the $b_n$ of the $N_f=2$ case: the number of cumulants (and derivatives)
involved in each $b_n$ grows as $n$ is increased. This is particularly obvious for odd $N_f$, where the sign of the determinant is involved,
which is a variable that the $N_f=2$ case knows nothing about.

Still, it is easy to see that a general relationship does exist for $b_2$:
\beq
b_2(N_f) = (N_f - 1) b_2 - (N_f - 2) b_2^{(0)},
\eeq
where $b_2$ is the coefficient for $N_f = 2$ and $b_2^{(0)}$ the coefficient for the noninteracting case.
We note the following limits are reproduced correctly by the above formula: $b_2(N_f=1) = b_2^{(0)}$, $b_2(N_f=2) = b_2$,
and $b_2(N_f) = b_2^{(0)}$ for all $N_f$ in the noninteracting limit. A more concise way to write this result is using
the noninteracting answer as a reference:
\beq
\Delta b_2(N_f) = (N_f - 1)\Delta b_2,
\eeq
where $\Delta b_2(N_f) = b_2(N_f) - b_2^{(0)}$ and $\Delta b_2 = b_2 - b_2^{(0)}$.

The relations among the virial coefficients derived in this section result from a double
expansion: the cumulant expansion of $\mathcal Z$ followed by the virial expansion,
which is a Taylor expansion on the fugacity $z$.
If the latter is replaced by an expansion with respect to a different
parameter (i.e., a different kind of source), then it may be possible
to generalize those relations to other quantities. Because $z$
enters in a special way, it is not \emph{a priori} obvious that such a
procedure applies to arbitrary quantities, however. We defer
further studies of such cases to future work.

%%%%%%%%%%%%%%%%%%%%%%%%%%%%%%%%%%%%%%%
\subsection{Empirical Fitting}

\begin{figure}[h]
\includegraphics[width=1.0\columnwidth]{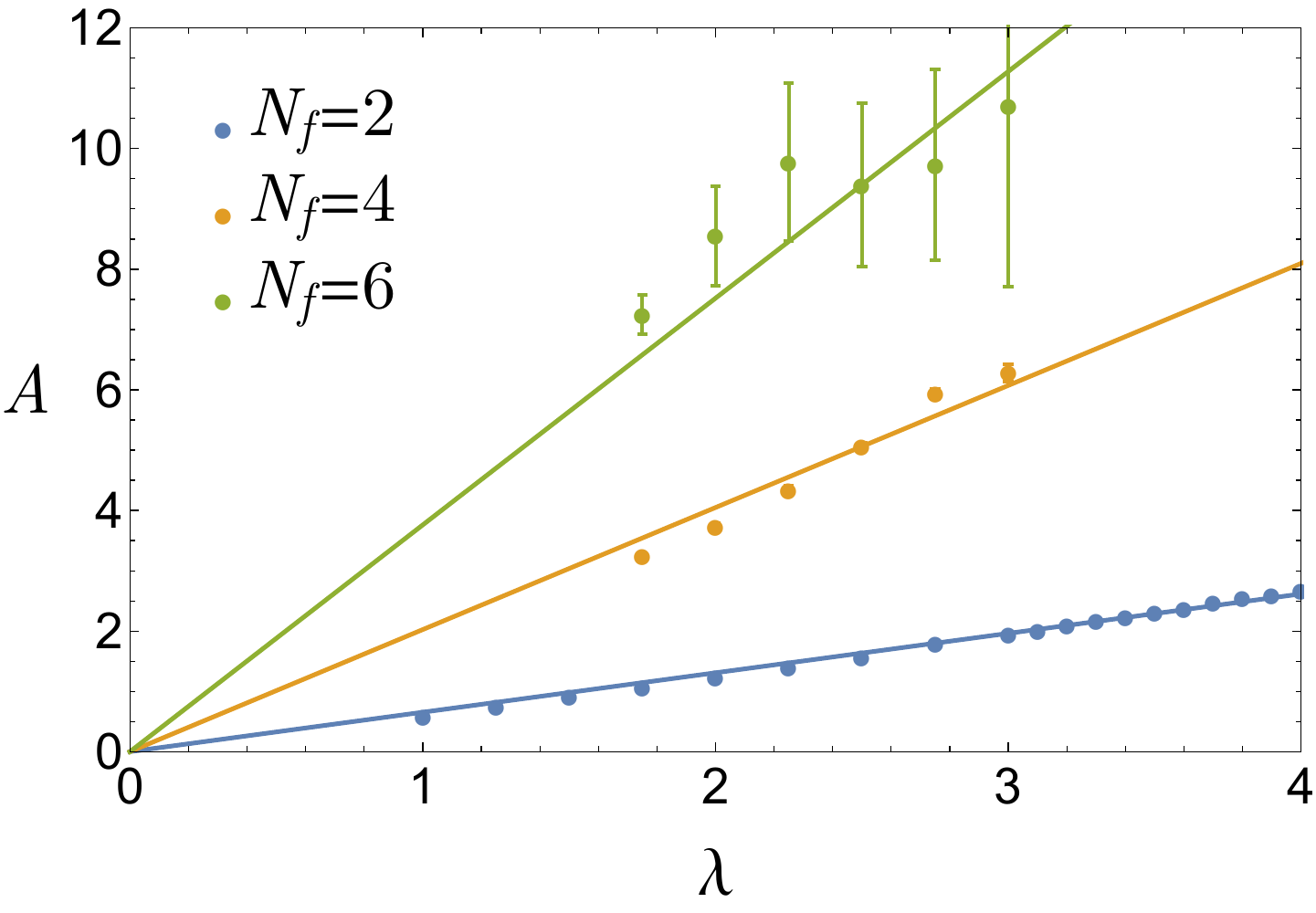}
\includegraphics[width=1.0\columnwidth]{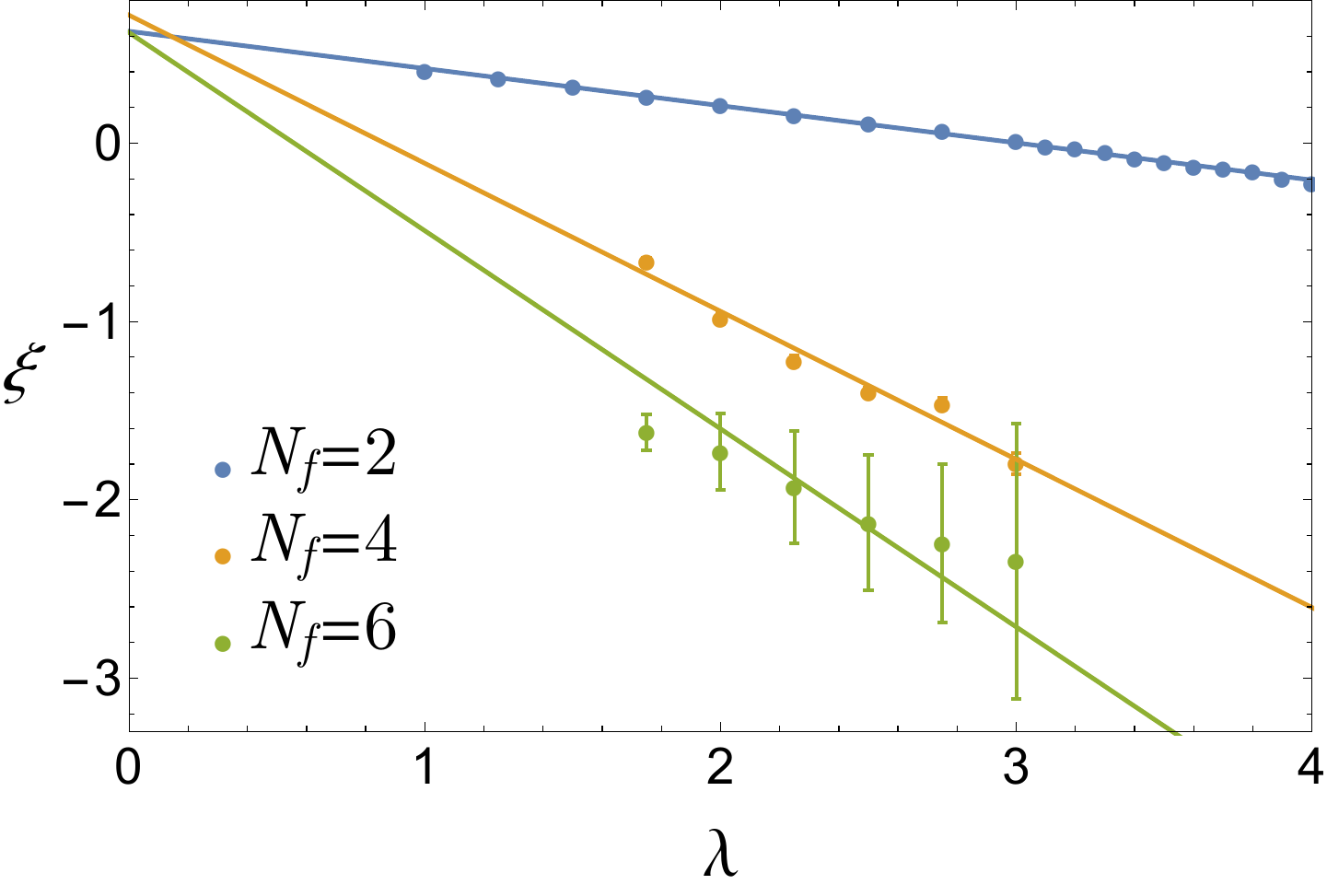}
\includegraphics[width=1.0\columnwidth]{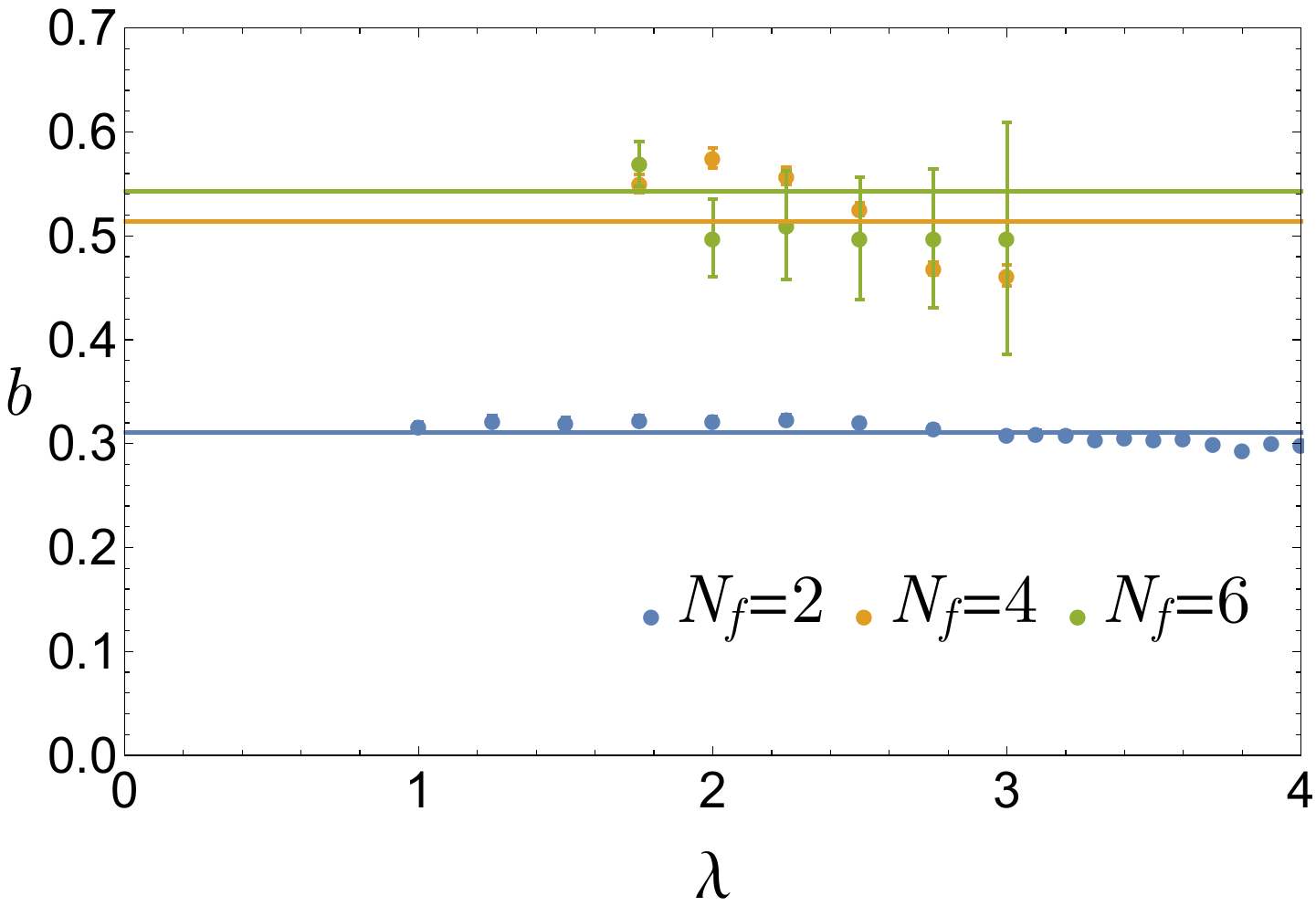}
\caption{\label{Fig:ampfig}(Color online) Fit parameter $A$ (top), $\xi$ (middle), and $b$ (bottom) of Eq.~(\ref{Eq:fit}) as a functions of the interaction
strength $\lambda$ for $N_f = 2,4,6$. The data points are the results obtained by fitting the Monte Carlo data; the solid lines are the fits to these data.}
\end{figure}

As seen in the plots shown above, increasing the number of flavors has a dramatic effect on the thermodynamics of the system,
which can be intuitively understood in terms of an enhanced interaction strength. The behavior of this 1D system is clearly
beyond the realm of perturbation theory and mean-field approaches. To encode our results in a useful form suitable for
future analyses, we develop empirical fits. These fits can be used to generate estimates outside the interaction strengths
examined here and for generating smooth curves underlying the data. The parametrizations were also performed for the data at
$N_f = 2$ in our previous work, Ref.~\cite{EoS1D}, for comparison. The model was generated according to the following formula:
\beq
\label{Eq:ansatz}
n/n_0(z) = \frac{n_0(\bar{z})}{n_0 (z)},
\eeq
where $z=\exp(\beta \mu)$ is the usual fugacity parameter and $\bar{z}$ is an {\it effective} fugacity whose form is set by taking
\beq
\label{Eq:fit}
\beta \mu \to \beta \mu + A \, (\mathrm{erf}(b \beta \mu - \xi) + 1),
\eeq
where $A$, $b$, and $\xi$ are fit parameters, and $\mathrm{erf}(x)$ is the error function; the shift by $+1$ was chosen to implement a smooth interpolation between
the non-interacting-type behavior at large negative $\beta\mu$ and the interacting form elsewhere. With this fit, the behavior of the interacting gas at
low fugacity tends to that of the noninteracting gas (i.e., $n/n_0 \to 1$ as $z \to 0$), while at large fugacity it reproduces the Pauli-blocked shape of the density distribution but with
a higher overall density due to the attractive interaction. The fit parameters as functions of the coupling $\lambda$ are shown in
Fig.~\ref{Fig:ampfig}.
The amplitude parameter $A$ must vanish as $\lambda \to 0$, which is consistent with Fig.~\ref{Fig:ampfig} (top), as that ensures the rescaled fugacity
will reproduce the noninteracting result in that limit. The amplitude $A$ varies linearly as a function of interaction strength, $A(\lambda) = a_A \lambda$, and the
coefficient $a_A$ itself varies linearly with $N_f$: $a_A(N_f) \simeq \alpha (N_f -1)$, where $\alpha = 0.73(3)$.
The shift parameter $\xi$ varies linearly with $\lambda$, $\xi(\lambda) = a_\xi \lambda + b_\xi$, as shown in Fig.~\ref{Fig:ampfig} (middle). The
coefficients $a_\xi$ and $b_\xi$ vary with $N_f$ as $a_{\xi}(N_f) \simeq 0.23(5) N_f - 0.2(2)$ and $b_\xi (N_f) \simeq 0.66(3)$.
The parameter $b$ does not vary significantly with interaction strength, as shown in Fig.~\ref{Fig:ampfig} (bottom).
%The parameter does not vary linearly with flavor, but can be fit with a hyperbolic tangent, $b(N_f) = -0.53[0.01] \tanh(1.33[0.05] - x)$ as a function of flavor.

Using these fits it is possible to interpolate between the curves generated using the Monte Carlo data. In addition, by integrating or taking derivatives it is
possible to generate functional estimates for the thermodynamic quantities presented in the paper. The particular choice of rescaling the fugacity
inside the Fermi-Dirac function for a noninteracting fermion gas seems robust and can be used to fit the density and pressure data for a two-dimensional
fermion gas presented in Ref.~\cite{Anderson:2015er}. Naturally, the physics underlying the specific shape of the density varies dramatically
with the spatial dimension. The proposed ansatz of Eqs.~(\ref{Eq:ansatz}) and (\ref{Eq:fit}) is based on the simple observation that density
distributions for fermions at finite temperature are typically smooth, monotonic interpolations between zero (at $\beta\mu \to -\infty$) and 1
(per flavor, at $\beta\mu \to \infty$).

%%%%%%%%%%%%%

%%%%%%%%%%%%%
%\begin{figure}[h]
%\caption{\label{Fig:scalefig}(Color online) The scale, $b$, of the fit in Eq.~(\ref{Eq:fit}) as a function of interaction strength $\lambda$ for $N_f = 2,4,6$. The data points are the shift values obtained by fitting the QMC data and the solid lines are the fits of this data.}
%\end{figure}
%%%%%%%%%%%%%%%%%%%%%%%%%%%%%%%%%%%%%%%

%%%%%%%%%%%%%%%%%%%%%%%%%%%%%%%%%%%%%%%%%%%%%%%%%%%%%%%%
\section{Summary and Conclusions}

We have performed a controlled, fully nonperturbative study of the thermodynamics of SU(4)- and SU(6)-symmetric
fermions with an attractive contact interaction. We report several quantities: density, pressure, compressibility, and Tan's contact.
We covered weakly and strongly coupled regimes as given by $3.0 \leq \lambda^2 \leq 9.0$,
as well as low to high fugacities as given by $-5.0 \leq \beta \mu \leq 8.0$. The latter covers the semiclassical regime $\beta \mu < -1.0$
as well as the deep quantum regime $\beta \mu > 1.0$. We employed lattice Monte Carlo methods that have been successfully
utilized before for similar studies, and discussed statistical and systematic uncertainties.

Our numerical results for the density equation of state show a behavior that is qualitatively similar to that of the SU(2) case but with
dramatic quantitative enhancement. The deviations from the noninteracting case are maximal for a $\lambda$-dependent value of
$\beta \mu$. As $z$ is increased from the semiclassical regime $z\ll1$, the strongly coupled regime is (roughly) accompanied by
the onset of quantum fluctuations as $\beta \mu = \ln z \simeq 0$ is approached.

One-dimensional Fermi systems with contact interactions are exactly solvable via the Bethe ansatz~\cite{BatchelorEtAl,TakahashiBook}.
This method, however, is restricted to uniform systems in the ground state (or close to it~\cite{BatchelorEtAlGaudinYang1,BatchelorEtAlGaudinYang2}).
Finite-temperature analyses require the thermodynamic Bethe ansatz, which involves solving an infinite tower of coupled nonlinear
integral equations~\cite{BatchelorEtAlGaudinYang1,BatchelorEtAlGaudinYang2}, which leads to potentially uncontrolled approximations \cite{HighSpin5,HighSpin6,HighSpin7,HighSpin8,HighSpin9,MaxwellContact1D,VignoloMinguzzi}.
The Monte Carlo techniques used here, on the other hand, have well-controlled systematic and statistical uncertainties.

In addition to our numerical answers, we used the auxiliary-field formulation of the quantum many-body problem to
show, in a general, nonperturbative fashion, that the virial coefficients of the SU($N_f$) case are fully determined by
the dynamics of the SU(2) problem.

Large-$N_f$ systems of the kind explored here were realized experimentally for the first time only 2 years ago~\cite{1DSUNExp}.
However, our results for the density and pressure equations of state, as well as the contact, are predictions for high-$N_f$ atomic
gases with attractive interactions.

%%%%%%%%%%%%%%%%%%%%%%%%%%%%%%%%%%%%%%%%%%%%%%%%%%%%%%%%
\acknowledgments
This material is based upon work supported by the
National Science Foundation under Grants No.
DGE{1144081} (Graduate Research Fellowship Program)
and
No. PHY{1452635} (Computational Physics Program).
\\

%%%%%%%%%%%%%%%%%%%%%%%%%%%%%%%%%%%%%%%%%%%%%%%%%%%%%%%%%%%
\appendix
\section{Systematics of the approach to the continuum limit \label{App:Systematics}}

\begin{figure}[]
\includegraphics[width=1.0\columnwidth]{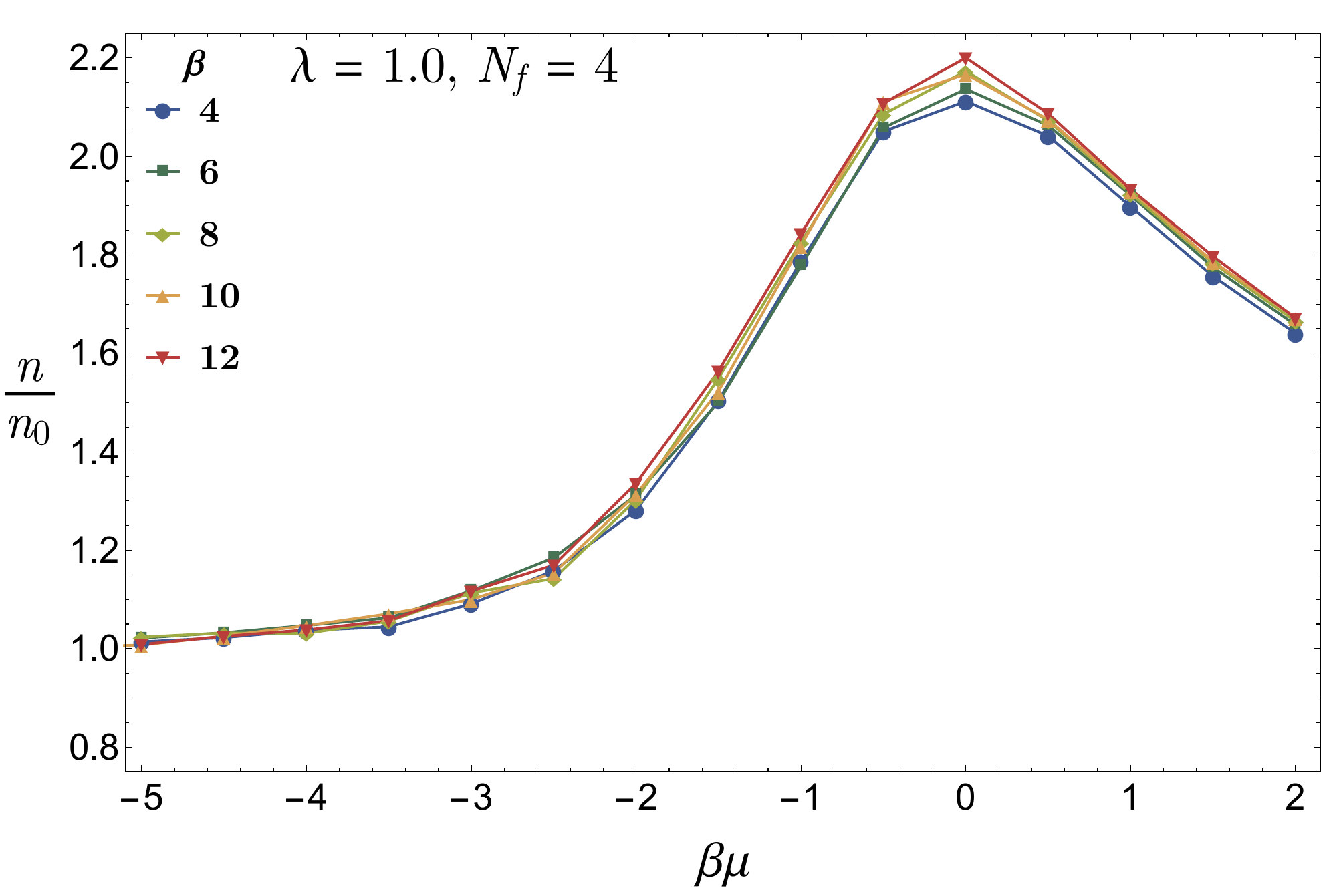}
\includegraphics[width=1.0\columnwidth]{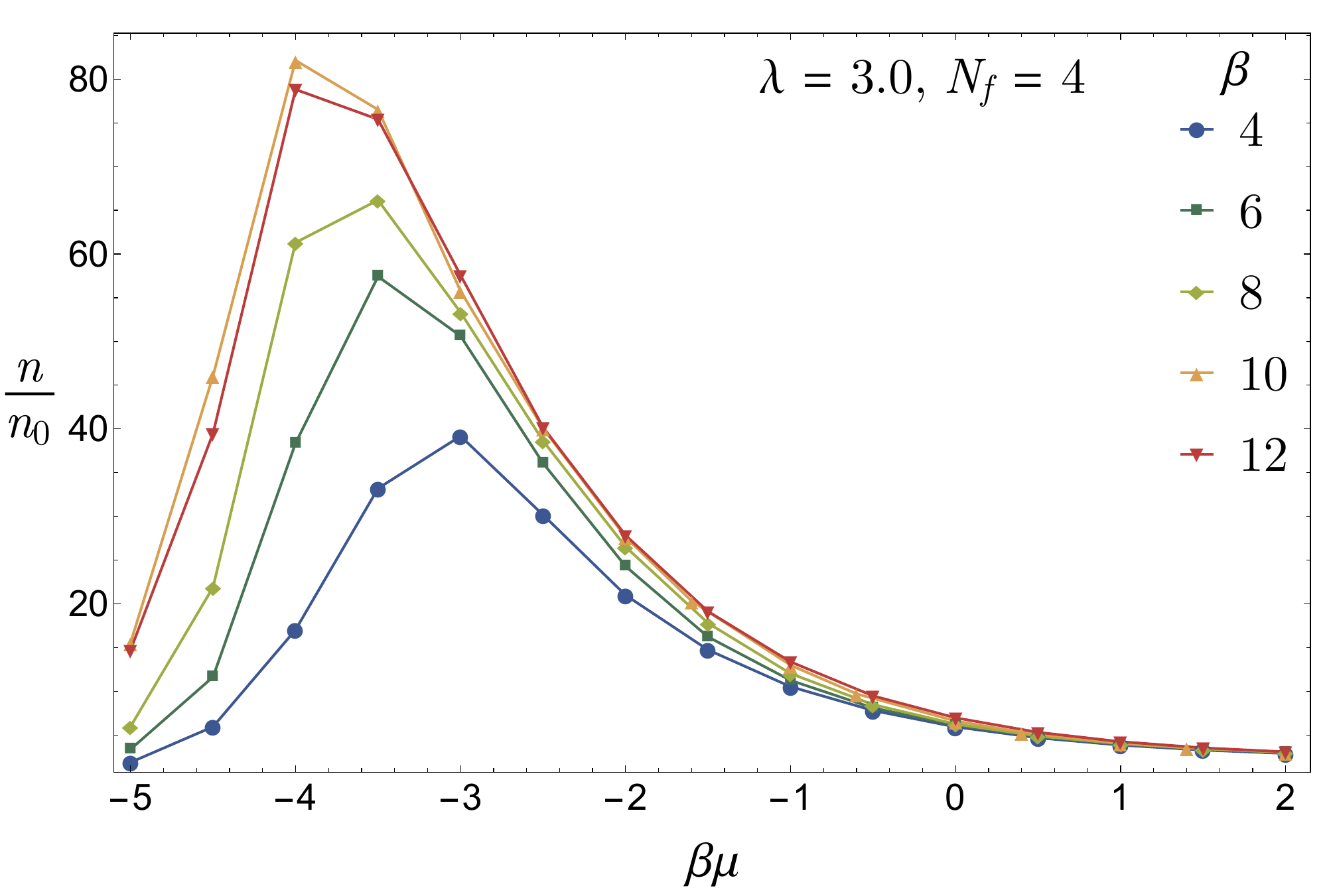}
\caption{\label{Fig:BetaSystematicsSC_nf4}(Color online) Density $n$ for $N_f=4$, in units of the noninteracting density $n_0^{}$,
as a function of $\beta\mu$ at weak coupling ($\lambda = 1.0$, top) and at the strongest coupling in this study ($\lambda = 3.0$, bottom),
for several values of $\beta$. Finite-$\beta$ effects are clearly visible, especially around the maximum.
Note the ranges in the $x$ and $y$ axes are different from those of Fig.~\ref{Fig:n_n0_nf4}.}
\end{figure}

\begin{figure}[]
\includegraphics[width=1.0\columnwidth]{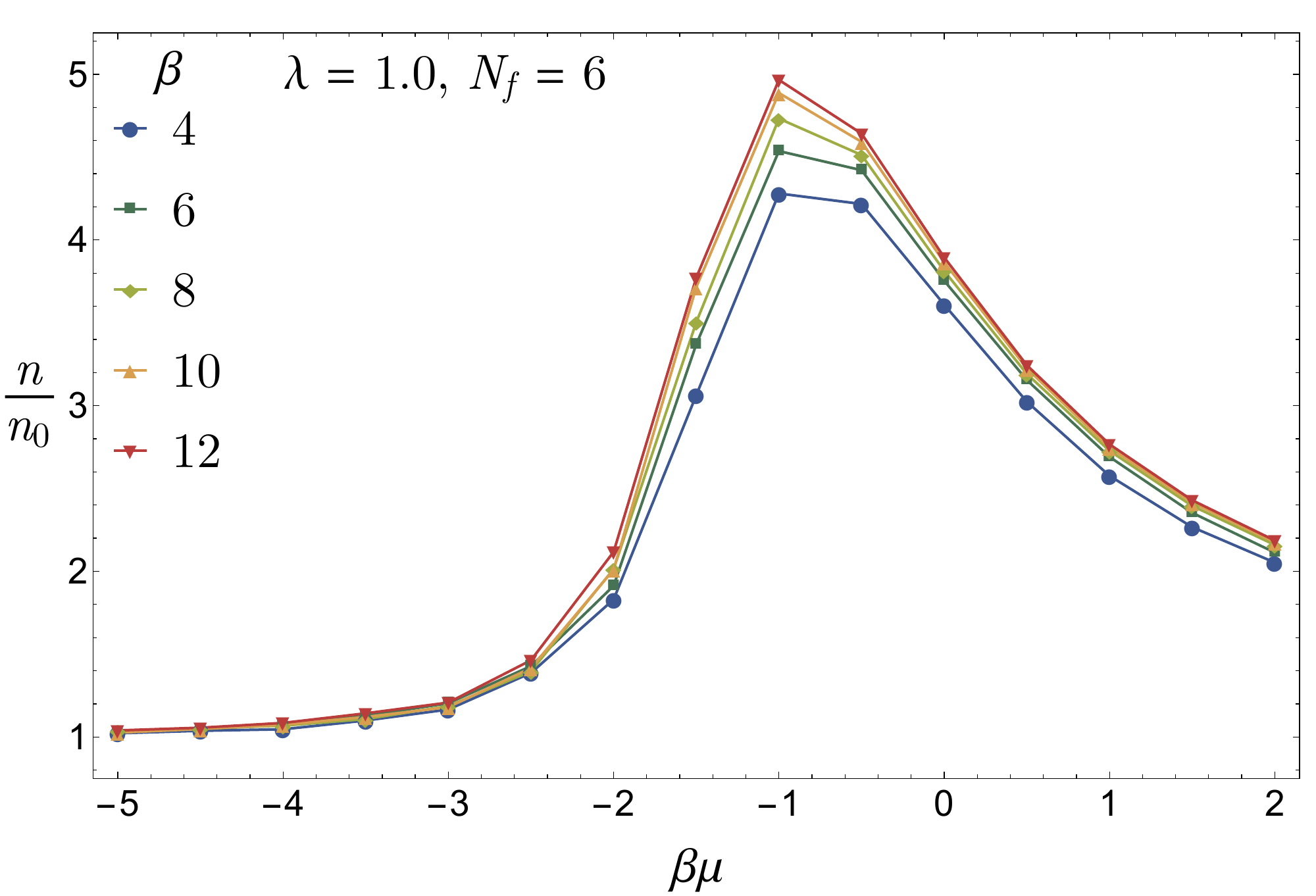}
\includegraphics[width=1.0\columnwidth]{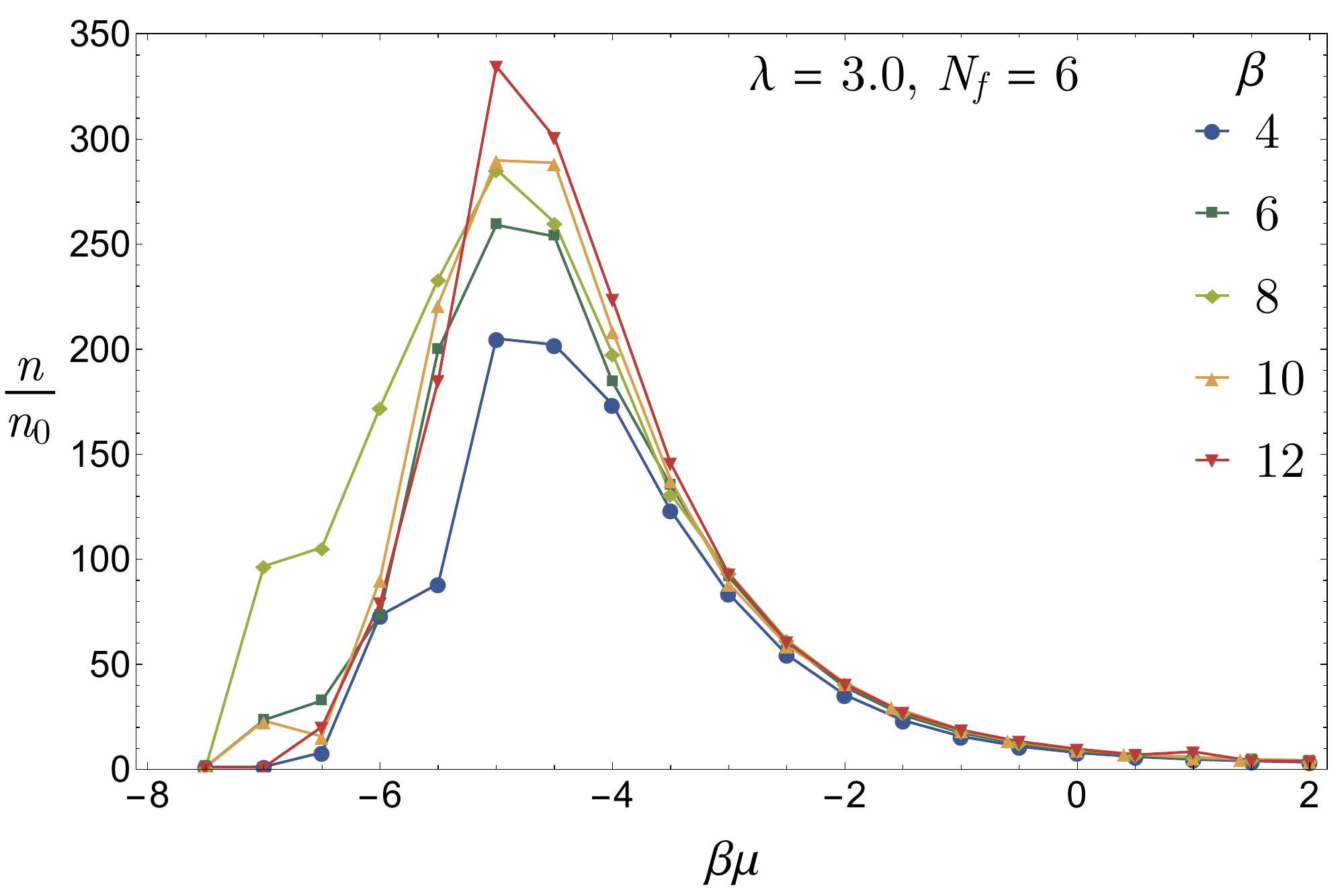}
\caption{\label{Fig:BetaSystematicsSC_nf6}(Color online) Density $n$ for $N_f=6$, in units of the noninteracting density $n_0^{}$,
as a function of $\beta\mu$ at weak coupling ($\lambda = 1.0$, top) and at the strongest coupling in this study ($\lambda = 3.0$, bottom),
for several values of $\beta$. Finite-$\beta$ effects are clearly visible, especially around the maximum.
Note the ranges in the $x$ and $y$ axes are different from those of Fig.~\ref{Fig:n_n0_nf4}.}
\end{figure}

In this section we report briefly on the systematic effects resulting from performing calculations at finite $\beta$.
As mentioned in the main text, the continuum limit is approached in our method when $\beta \to \infty$, and
different quantities approach their limit at different rates, which also depend on the values of other input parameters
(e.g., $\beta \mu$). As we show in Figs.~\ref{Fig:BetaSystematicsSC_nf4} and~\ref{Fig:BetaSystematicsSC_nf6}, the
convergence to the large-$\beta$ limit is not uniform: where the interaction and quantum effects dominate, the convergence properties are
poorer. This is clearer at strong coupling (Fig.~\ref{Fig:BetaSystematicsSC_nf4}, bottom) than at weak coupling (Fig.~\ref{Fig:BetaSystematicsSC_nf4}, top);
indeed, the latter is essentially converged already at $\beta = 4$, whereas the former still shows finite-$\beta$ effects
even at $\beta=8$ in some regions. From these graphs, we infer that the largest systematic uncertainties
due to finite $\beta$ are on the order of $10\%$ in the worst-case scenario. We stress that that is an upper bound for these systematic effects.
Those effects are most prominent around the maximum in $n/n^{}_0$; they are apparent for the strongest couplings
we have studied ($\lambda=3$) and are small for weak coupling ($\lambda=1$).

%%%%%%%%%%%%%
%%%%%%%%%%%%%
%%%%%%%%%%%%%
%%%%%%%%%%%%%

%%%%%%%%%%%%%%%%%%%%%%%%%%%%%%%%%%%%%
\section{Derivation of partition function formula for $N_f$ flavors \label{App:Derivation}}

In this section we provide a schematic derivation of the form of the $N_f$-flavor partition function in terms
of a field integral. The starting point is the definition
\beq
\mathcal Z = \text{Tr}\left[e^{-\beta(\hat H - \mu \hat N)} \right],
\eeq
where we assume for this derivation that $\mu$ is the same for all fermion species [as befits the SU($N_f$)-symmetric case] and that
the interaction is pairwise among all flavor pairs, such that, writing $\hat H = \hat T + \hat V$, the interaction is
\beq
\hat V = - g \int dx \; \hat n_1 \hat n_2 - g \int dx \; \hat n_2 \hat n_3 + \cdots ,
\eeq
where we have labeled the flavors as $1,2,3,\dots,N_f$ and the dots include all possible flavor pairs.
Upon a Trotter-Suzuki factorization (see, e.g., Ref.~\cite{Drut:2012md}), we are left with the task of considering, at each point in space,
\bea
&&\exp\left( \tau g \, \hat n_1 \hat n_2 + \tau g \, \hat n_2 \hat n_3 + \cdots \right) = \\
&&\ \ \ \ \ 1 + A^2(\hat n_1 \hat n_2 + \hat n_2 \hat n_3 + \cdots) + \mathcal O(A^4) ,
\eea
where $A^2 = e^{\tau g} - 1$, again the dots include all possible flavor pairs, and we have also used the exact property
\beq
\exp\left( \tau g \, \hat n_1 \hat n_2 \right) = 1 + A^2 \hat n_1 \hat n_2
\eeq
for each pair of fermion flavors appearing in the interaction. Note that the size of the subleading terms $\mathcal O(A^{2n})$
are controlled by the size of $\tau g$ and vanish as $(\tau g)^{n}$ when $\tau g \to 0$.

We next notice that a single Hubbard-Stratonovich transformation is able to reproduce the leading terms written above.
Indeed, one could use for instance the following discrete form:
\bea
&& \frac{1}{2}\sum_{\sigma=\pm 1} (1 + A \sigma \hat n_1)(1 + A\sigma \hat n_2)\cdots(1 + A\sigma \hat n_{N_f}) = \nonumber \\
&& \ \ 1 + A^2(\hat n_1 \hat n_2 + \hat n_2 \hat n_3 + \cdots) + \mathcal O(A^4).
\eea

Thus, within the above approximation a single Hubbard-Stratonovich field is enough to factorize the interaction. Note that the approximation is already present in the use of the Trotter-Suzuki factorization, such that no new approximations
are actually being introduced. Each factor on the right-hand side of the above equation is a one-body operator that
affects only one of the fermion flavors.

From this point on, the usual derivation (see, e.g., Ref.~\cite{Drut:2012md}) proceeds normally
and one may ``integrate out'' the fermions to produce a fermion determinant for each species. As all of the fermion species are identical,
one obtains the same determinant for each of them, which yields the result advertised above, namely, that
the generalization of the $N_f = 2$ case to $N_f$ identical species only requires replacing the power of 2 in the determinant
with a power of $N_f$.

We stress that this derivation is simply one way to arrive at the standard expressions used in this work for arbitrary $N_f$.
The analogs of such standard expressions are used for electrons throughout condensed matter as well as for gluons
in quantum chromodynamics and are therefore not new.

%%%%%%%%%%%%%%%%%%%%%%%%%%%%%%%%%%%%%%%%%%%%%%%%%%%%%%%%%%%

%%%%%%%%%%%%%%%%%%%%%%%%%%%
\end{document}